%Paper: hep-ph/9206201
%From: SHAHAR@slacvm.slac.stanford.edu
%Date: Mon, 01 Jun 1992 04:13 -0800 (PST)

\input phyzzx
\doublespace
 \newtoks\slashfraction
 \slashfraction={.13}
\def\diag#1#2#3{\left( \matrix{#1&\ &0\cr
                               \ &#2&\ \cr
                               0&\ &#3\cr} \right) }
\def\vector#1#2#3{\left( \matrix{#1\cr #2\cr #3\cr} \right) }
\def\lm{\lambda}
\def\lf{\lbrack}
\def\rr{\rbrack}
\def\prl{{\scriptscriptstyle \parallel}}

\def\etp{{\scriptscriptstyle (\eta)}}
\def\pls{{\scriptscriptstyle (+)}}
\def\zro{{\scriptscriptstyle (0)}}
\def\mns{{\scriptscriptstyle (-)}}
\def\zro{{\scriptscriptstyle (0)}}
\def\Isp{{\scriptscriptstyle (I)}}
\def\Jsp{{\scriptscriptstyle (J)}}
\def\Ksp{{\scriptscriptstyle (K)}}
\def\Is{{\scriptscriptstyle I}}
\def\Js{{\scriptscriptstyle J}}
\def\Ks{{\scriptscriptstyle K}}
\def\Asp{{\scriptscriptstyle (A)}}
\def\As{{\scriptscriptstyle A}}
\def\Bs{{\scriptscriptstyle B}}
\def\Cs{{\scriptscriptstyle C}}
\def\Hs{{\scriptscriptstyle H}}

\def\IspT{{\scriptscriptstyle (I)T}}
\def\supT{{\scriptscriptstyle T}}

\def\bolt{{e^{\scriptscriptstyle -E_i(\vec{k})/T}}}

\def\cala{{\cal A}}

\def\cl{{\cal L}}
\def\calc{{\cal C}}
\def\calt{{\cal T}}
\def\prt{\partial}
\def\pr{\prime}

\def\kp{\kappa}
\def\om{\omega}
\def\al{\alpha}
\def\eps{\epsilon}
\def\uem{{\rm U(1)_{em}}}
\def\zt{{\rm{\bf Z}_2}}

\def\ot{{\rm O(2)}}
\def\sod{{\rm SO(3)}}
\def\soten{{\rm SO(10)}}

\pubnum={SLAC-PUB-5828}
\pubtype={T}
 \date{May 1992}
\titlepage
\title{Baryogenesis from Unstable Domain Walls}
\foot{Supported by the Department of Energy, contracts
 DE-AC03-76SF00515 and DE-FG03-84ER-40168.}
\author{Shahar Ben-Menahem}
\address{Stanford Linear Accelerator Center\break
Stanford, California 94309}
\author{Adrian R.~Cooper}
\address{Department of Physics\break
University of Southern California\break
Los Angeles, California 90089-0484}
\medskip
\abstract
There exists a class of cosmic strings that turn matter into antimatter
(Alice strings).
In a GUT where the unbroken gauge group contains charge conjugation
($C$), such strings form when a phase transition
 renders $C$ a discrete symmetry.
They become boundaries of domain walls at a later, $C$-breaking
transition. These `Alice walls' are cosmologically harmless, but can
play an important role
 in baryogenesis. We present a three-generation toy model
with scalar baryons, where a quasi-static Alice wall (or a gas of such
walls) temporarily gives rise to net baryogenesis of uniform sign
   everywhere in space. This becomes a permanent baryon excess if
 the wall shrinks away early enough.
 We comment on the possible relevance of a similar mechanism to
 baryogenesis in a realistic $\soten$ unification model, where
 Alice walls would form at the scale of left-right symmetry breaking.
\submit{Nuclear Physics \bf B}
\endpage
\chapter{Introduction}
\REFS\schwarz{A.S.~Schwarz \journal Nucl.Phys.&B208 (82) 141; \hfil \break
A.S.~Schwarz and Y.S.~Tyupin \journal Nucl.Phys.&B209 (82) 427.\hfil }
\REFSCON\preskill{J.~Preskill and L.~Krauss \journal Nucl.Phys.&B341
(90) 50; \hfil \break
M.~Bucher, H-K.~Lo, and J.~Preskill, ``Topological Approach
to Alice Electrodynamics,'' Caltech Preprint CALT-68-1752 (1992).\hfil }
\REFSCON\coleman{M.~Alford, K.~Benson, S.~Coleman, J.~March-Russell,
and F.~Wilczek \journal Nucl.Phys.&B349 (91) 414.\hfill }
\REFSCON\brek{L.~Brekke, W.~Fischler, and T.~Imbo,
 ``Alice Strings, Magnetic
Monopoles, and Charge Quantization,'' Harvard Preprint
 HUTP-91/A042 (1991).}
\REFSCON\preswise{J.~Preskill, S.P.~Trivedi, F.~Wilczek, and M.B.~Wise
\journal Nucl.Phys.&B363 (91) 207.}
\REFSCON\wallnet{T. Kibble, G. Lazarides and Q. Shafi
\journal Phys.Lett.&113B(82)237;\nextline T. Kibble, G. Lazarides
 and Q. Shafi
\journal Phys.Rev.&D26(82)435;\nextline
A. Vilenkin and A.E. Everett
\journal Phys.Rev.Lett.&48(82)1867;\nextline
see also\ \lf\preswise\rr\   and references therein.}
\REFSCON\dtmon{
S.~Ben-Menahem and A.R.~Cooper, SLAC-PUB-5805.}
\REFSCON\scond{
S.~Ben-Menahem and A.R.~Cooper, in preparation.}
\REFSCON\dine{
M. Dine, R.G. Leigh, P. Huet, A. Linde and D. Linde, SLAC-PUB-5741,
 SCIPP-92-07, SU-ITP-92-7, and references therein.}
\refsend
Alice strings [\schwarz,\preskill,\coleman] are a class
 of cosmic strings
with the remarkable property that a particle traveling around one will
come back as its own antiparticle.  In a Grand Unified Theory in which
the charge conjugation operator $C$ is contained in the original gauge
group, such strings form via the Kibble mechanism when the symmetry is
 broken to a smaller group having $C$ as a discrete (i.e. not deformable
 to $1$) symmetry. This occurs, for example, for certain $\soten$ GUTs.
 \par
In the vacuum we inhabit, of course, $C$ is not a symmetry. This means
it must have been spontaneously broken at some later phase transition.
At that epoch, the Alice strings would have become boundaries of domain
 walls. We shall henceforth refer to the latter as `Alice walls'.
The dynamics of cosmological networks of string-bounded walls has been
studied \lf\wallnet\rr. The walls eventually shrink via surface tension,
string intercommutation and nucleation of new string loops. Thus they
never dominate the energy density of the Universe, and can have
interesting cosmological effects while they last. \par
 Until recently it was believed that Alice strings could only have
existed in pre-inflationary epochs, since a phase transition
 giving rise to Alice strings also tends to form magnetic monopoles in
 abundances ruled out by experiment\ \lf\brek\rr  .
  However, in ref.\dtmon\  we
  gave an example of a natural symmetry-breaking scheme in which
 this problem is avoided. The models considered in ref.\dtmon\  were
  variants
 of the original Alice-string model of ref.1 in which an $\sod$
 gauge symmetry is spontaneously broken to $\ot$, and then (optionally)
 to smaller groups. \par
  The present paper is dedicated to the study, in such a toy
 model, of a novel baryogenesis mechanism involving Alice walls.
 The original gauge group is $\sod$, and the
 (scalar) baryons belong to $\sod$ triplets. We define the
  {\it baryon
 number}, $B$, to be the electric charge for members of these `baryonic
 triplets', and zero for all other fields.
 Here `electric charge' refers to the $\uem$ subgroup of
$\sod$ generated by $T_3$. Baryon number is violated by perturbative
 processes, such as exchange of
 the charged\foot{By `charged' we shall always mean $\uem$ charge.}
  vector bosons $W_\pm$ of $\sod$, although no net baryogenesis can
   occur while $C\equiv\exp(\pi iT_2)$ is still conserved. When
 $\sod$ is broken to  $\ot$ (which is the
 semidirect product $\uem\times\{1,C\}$), the
 $W_\pm$ bosons become massive and the perturbative violations of
 baryon number become suppressed as the temperature continues to fall.
 However, this phase transition also creates Alice strings, through
 the Kibble mechanism. In the presence of such a string $C$ is
 still conserved (it is an element of $\ot$),
  but $B$ is globally ill-defined since when a `baryon'
 is taken once around a string it becomes an `antibaryon'. Locally,
 however, $B$ may be sensibly defined and is still weakly violated
 by perturbative processes \foot{We thank J. Preskill for a discussion
  concerning this point.}.\par
At a lower temperature, we spontaneously break $C$ by allowing
a (non-baryonic) $\sod$ Higgs triplet to acquire a VEV
 \foot{Parity is conserved in our model, and so $CP$ and $C$ are not
 independent symmetries. This circumstance will of course change in
 the realistic $\soten$ model.}.
 A closed Alice string loop now becomes the boundary of
  an Alice wall, where the triplet Higgs develops a kink.
{\it Outside} such a wall, it is clear that $B$ may be globally defined.
In this case, the passage of a baryon through the wall will be seen as
a $B$-violating process --- it will emerge on the other side as an
antibaryon (and vice versa). If we could arrange for baryons to bounce
 off the wall more frequently than antibaryons, say, then we would have a
 means of driving baryogenesis. In practice, we find that the model
 must be complicated slightly to achieve this: it must contain at least
three copies (`generations') of baryonic triplets, and also at least
three Higgs triplets. The transmission rates of baryons and antibaryons
through the wall are generically different in the extended model.\par
 If we now impose that (prior to wall formation) the ambient
 baryon-antibaryon plasma has $B=0$ and is out of thermal equilibrium,
then baryogenesis will {\rm initially} occur. This result holds even for
 a static Alice wall --- a novel result so far as we are aware\foot{Note
 that even in this case Sakharov's conditions are met since the ambient
 plasma is not in thermal equilibrium.}. In a cosmological setting there
 will be a gas of such Alice walls, each driving baryogenesis. It is
 important to stress that the {\it sign} of the baryon excess will be
 uniform throughout space. However, if the walls are left intact
 indefinitely, $B$ will execute damped oscillations and return
 asymptotically to zero. We must therefore allow the walls to
 shrink away sufficiently quickly that a net $B$ remains.
\par
This baryogenesis mechanism occurs at the tree level. It is completely
classical if we think of the scalar baryons as wave packets. It is more
 useful, however, to think of the relevant transmission
 (and reflection) rates as squares of moduli of quantum
 amplitudes, in a formalism where the baryons are first-quantized and
 evolve in the classical background of the Alice wall. \par
  In our specific model, net baryon number implies also a net electric
 charge. Thus the Alice wall must accumulate charge (Cheshire or
otherwise) to offset the baryon number it creates around it;
 in the parameter regime we shall focus on, only Cheshire (i.e.
 nonlocalized) charge accumulates. This charge continuously decays as
 it forms, by electrically polarizing the
   surrounding plasma and also by the quantum-mechanical
  emission of charged particles. Since the plasma is composed of
 other charged particles besides (anti)baryons, the decay of
 Cheshire charge will only partially cancel the accumulated $B$, not
 eliminate it.
 \par
 The Alice walls in our model
 may or may not exhibit the Meissner effect.  This depends on whether the
charged Higgs components acquire VEVs in the wall core.
We refer to these two cases as the
  {\it superconducting} and {\it normal} cases, respectively.
In the version where we
 find baryogenesis, the normal phase is chosen for all three Higgs
  triplets, since this is simplest. The case of superconducting walls,
  however, has its own points of interest, quite apart from baryogenesis.
  We shall return to this case in a future publication\ \lf\scond\rr.
\par
  So far we have been discussing a toy model. Some viable GUT schemes
  can also form Alice strings and walls; we wish to raise the
 question of whether a variant of our mechanism can play an important
 role in baryogenesis for such GUT's---e.g. for the group $\soten$. This
 question is currently under investigation; we
  discuss it briefly in the concluding section.  \par
 The remainder of the paper is organized as follows. In section 2
 we set up the toy model. We specialize to a Higgs potential
 such that both an Alice string and the wall it bounds
 are formed, and such
 that the wall is normal (non-superconducting). In this regime, the wall
 is described as kinks in the three neutral Higgs fields.
 In section 3, we analyze the spectrum and discrete
  symmetries of the model; it is seen that
 $C$ (and thus $CP$) are broken in the space surrounding the
 wall. Section 4 is the main part of the paper. It is shown there
  that the discrete symmetries are not capable of
 constraining the baryon and antibaryon transmission rates sufficiently
 to rule out baryogenesis. A parameter regime is chosen
 which simplifies the physics. In particular, the
  heaviest among the three mass-eigenstate baryons is decoupled.
 We next proceed to derive the momentum- and
 position-averaged rate equations for the concentrations of the two
 lightest baryon species, and their antibaryons
 \footnote{\flat}{The fact that we
 are able to completely decouple one of the three mass-eigenstate
 baryon species yet still obtain
 baryogenesis, does not contradict our previous statement that at
 least three generations are needed. This is explained in section 4.}.
 The asymmetry between the transmission rates of baryons
 and antibaryons through the wall is computed in the Born approximation.
 It is thus shown that an
  out-of-equilibrium plasma with $B=0$ initially accumulates a net
 baryon number, which becomes permanent if the wall disappears in time.
  This baryon number has a
 uniform, physically-meaningful sign throughout space. This
 remains true in the presence of a gas of such walls. In section 5 we
 summarize our conclusions, and briefly discuss ongoing work on
  baryogenesis in $\soten$.
\bigskip
\chapter{The Model}
Our model is an $\sod$ gauge theory with the following matter content:
 An isospin-2 Higgs field\footnote{\dag}{`Isospin' refers to the
  gauge-group transformation properties.}
  $\Phi_{ab}$ (a $3\times 3$ traceless symmetric matrix);
 three scalar `baryonic triplets' $\psi_a^{(\Is)}$, with
  $I$ ranging from $1$ to $3$; and three Higgs triplets, $v_a^{\Isp}$.
 We shall refer to $I$,$J$, etc. as {\it generation} indices.
 The lower-case $a$,$b$ etc. will denote gauge indices; repeated gauge or
 Lorentz indices will be summed over except when stated
 otherwise, but not generation indices. We employ the West-Coast
 metric $(1,-1,-1,-1)$. We also include one additional nonbaryonic
 isotriplet, $u_a$; its sole interaction is a minimal coupling to the
 gauge fields. The only role $u$ will play in the model is to cancel
 the Alice wall's accumulated electric Cheshire charge. We will not
 need to explicitly write the $u$-dependent terms in the action.\par
 \par The Lagrangian density is as follows:
 $$\left.\eqalign{\cl=&-{1\over{4e^2}}F^a_{\mu\nu}F^{a\mu\nu}+
 {1\over 2}tr(D_\mu\Phi)^2
 +{1\over 2}\sum_\Is(D_\mu v^{\Isp})^2+
 {1\over 2}\sum_\Is(D_\mu\psi^{\Isp})^2\cr
 &-V_1(\Phi)-\sum_\Is V_2^{\Isp}(v^\Isp,\Phi)-V_3(v)
 -V_4(\psi,\Phi)-V_5(\psi,v)}\right.\eqno(2.1)$$
 with (the superscript $T$ denotes isovector transposition):
 $$\left.\eqalign{V_2^{\Isp}(v,\Phi)=&\lm_\Is v^{\supT}\Phi v
 +a_\Is v^2\cr &+\rho_\Is(v^2)^2}\right.\eqno(2.2)$$
 $$V_3(v)=-\eps\sum_{\scriptscriptstyle I<J}v^{\Isp}
 \cdot v^{\Jsp}\eqno(2.3)$$
$$V_4(\psi,\Phi)={1\over 2}\sum_{\Is\Js}
\nu_{\Is\Js}\psi^{\IspT}\psi^{\Jsp}
 -{\kp\over 2}\sum_\Is \psi^{\IspT}\Phi\psi^{\Isp}\eqno(2.4)$$
$$V_5(\psi,v)=-{b\over 2}\sum_{\Is\Js\Ks}\eps_{\Is\Js\Ks}
\eps_{abc}\psi^{\Isp}_a
 \psi^{\Jsp}_b v^{\Ksp}_c\eqno(2.5)$$
 where $e$ is the gauge coupling, and $\lm_\Is$,$\rho_\Is$,$\eps$,$\kp$
and $b$ are all chosen to be positive. The gauge fields are $A_\mu^a$,
and their field strengths $F_{\mu\nu}^a$.
\par
We now explain the motivation of the various potential terms.
 $V_1$ is some polynomial in traces of powers of
  the matrix $\Phi$, chosen
  to have a minimum (unique up to a gauge transformation) at
 $$\Phi=\al\diag 1 1 {-2},\quad \al>0\,;\eqno(2.6)$$
 we shall not require a specific form for $V_1$ here. The role of the
 $V_2^{\Isp}$ terms
 is to give the triplet Higgs fields vacuum expectation
  values (VEVs); apart
  from the standard Mexican-hat potentials, they contain alignment terms
 (with couplings $\lm_\Is$) which tend to
  align $\vev{v_a^{\Isp}}$, in
 isospin space, with respect to $\Phi$. This alignment is required
 to preserve the $\uem$ symmetry, as explained below.
The neutral members of the Higgs triplets
  must acquire VEVs in order to break $C$. As we shall see, these VEVs
 are constant in the bulk of space, and have kinks at the wall since
 the Alice twist requires them to vanish on some surface within
  the wall.\par
 The Higgs couplings are generation-dependent, and in a cosmological
 setting they are also temperature- (and thus time-) dependent. Thus,
  $a_\Is$ should decrease with temperature, and at some epoch they
 will be in a regime such that all
  the Higgs triplets acquire VEV's, and $C$ is broken.
 \par
  The signs of $\vev{v_3^{\Isp}}$ for different
   $I$ are {\it a priori} uncorrelated.
 The purpose of the term $V_3$ is to correlate them, thus ensuring that
 there are only two true vacua after $C$ is broken---and as
 one is the $C$-reflection of the other, these are gauge equivalent.
  The parameter $\eps$ can be chosen as small
 as desired. Together with $V_2$, it constitutes the mass-squared
 matrix of the $v$-fields in the $C$-symmetric vacuum. The $V_3$
 term correlates not only the bulk VEVs of the Higgs triplets, but also
 the locations of the kinks for different values of $I$.
 \par
 The potential $V_4$ contains the $\psi$ mass term and a $\psi\psi\Phi$
 Yukawa interaction. The latter is introduced in order to break the
 degeneracy between the charged and neutral members of the baryonic
 multiplets. By (2.4),(2.6) we have
$$m_\As^2=\mu_\As^2-\kp\al,\quad M_\As^2=\mu_\As^2+2\kp\al
\quad\;(A=1,2,3)\eqno(2.7)$$
where $\mu_\As^2$ are the eigenvalues of the symmetric matrix
 $\nu_{\Is\Js}$, and
 $m_\As$,$M_\As$ are the masses of the charged
  and neutral $\psi$ fields,
 respectively, before $C$ is broken. This mass splitting between
 charged and neutral baryons is not really
 necessary, but it is convenient, since we can now proceed to make
 the neutrals very massive:
 $$M_\As=O(\al),\quad m_\As\ll\al \eqno(2.8)$$
 where $\al$ is the scale of the $\Phi$ VEV.
 This is accomplished by choosing $\mu_\As=O(\al)$, $\vert\mu_\As-
\mu_\Bs\vert\ll\al$ and tuning $\kp$ to produce a hierarchy
 between $M_\As$, which are at the $\sod$-breaking scale,
  and $m_\As$. Thus neutral baryonic particles can be assumed to
  have already decoupled at the temperatures where the
 Alice wall forms. This suits our purpose, since these particles have
 no baryon number---we wish to concentrate on
  the baryons and antibaryons.
\par
 Finally, there is the potential $V_5$. This consists of Yukawa couplings
 of the form $\psi\psi v$, and serves to couple the baryons to the wall.
 In the bulk (outside the wall), it  changes the mass-squared
 matrix of the baryons in the $C$-broken phase, so the bulk
  definition of the three baryons (and their antibaryons)
  changes. In this phase a mass-eigenstate baryon is, in general,
 no longer in the same gauge multiplet as its $CPT$-conjugated
 antiparticle. Note that despite the breaking of
   $C$ and $CP$, each baryon-antibaryon pair is still
 degenerate, as required by the $CPT$ theorem.
  In addition to its effect in the bulk, we shall see that the
   $V_5$ potential gives rise to different transmission probabilities
 for incoming baryons and antibaryons of the same energy. Since each
 transmitted baryon (antibaryon) is turned into an antibaryon (baryon)
\footnote{\diamondsuit}{In general a baryon is sometimes
 transmitted while staying
 a baryon. However, we shall choose the parameters of the model so this
 does not happen. This choice will also ensure that a {\it reflected}
 baryon (antibaryon) remains unflipped.}, this
  asymmetry enables baryogenesis. This will be explained
 in detail in section 4. \par
 The VEV eq.(2.6)
  breaks the gauge group $\sod$ down to $\ot$,
 which has two connected components and is the semidirect
 product of $\uem$ and $\zt\equiv\{1,C\}$. The group
 $\uem$ consists of isospin
 rotations about the internal $3$-axis; $C$ is charge conjugation, and
 we choose it to be a rotation by $\pi$ about the internal $2$-axis.
  Stable Alice strings exist in the $\ot$ phase. We wish to consider
 processes in the background of a closed Alice string,
  so we must begin by twisting all charged fields by $C$ around curves
  linking the Alice-string loop \foot{We shall not need
 to consider the structure of the string core in this paper. Since
 we assume a hierarchy between the two symmetry-breaking scales, it
 is natural for the wall thickness (and of course its size) to be
 much larger than the string core. We {\it shall}, however, require
  details of the kink configuration at the core of the Alice
 {\it wall}.}.
  Thus, the physical gauge defined by eq.(2.6) is valid in all of
 space excluding some branch-cut surface. This surface
  is arbitrary, except
 that its boundary must be the string. Any given field is matched to
 itself across the cut via the $C$ operation. Thus
  $\vev{v_3^{\Isp}}$ flip their signs at the cut, $\vev{v_2^{\Isp}}$
 are continuous, etc.
 \par It is straightforward to check that the $vv\Phi$ alignment term
 in $V_2$ forces $\vev{v^{\Isp}}$ in the bulk to lie along the internal
 $3$ direction. The term was chosen for precisely that reason: we wish
 to break $C$, but not $\uem$, in the bulk of space
\foot{Except during a brief Langacker-Pi phase, which is desirable
  in order to get rid of magnetic monopoles\ \lf
 \dtmon\rr. Such a phase does not arise in the model described here,
 but we shall see below that there is a parameter range
 for which the kink that defines the wall is superconducting.
 In this regime, $\uem$ is broken {\it in the wall}, though it
 is still a good symmetry in the bulk. However, in this paper we
 concentrate on the regime where $\vev{v_a^{\Isp}}=0$ everywhere for
 $a=1,2$.}.
\par We next investigate the structure of a static Alice wall.
  We shall assume the string loop to be a smooth curve
 (e.g. a circle), statically held in place, with the pancake-shaped
  wall centered at the planar disk bounded by the string.
 \foot{The wall
 is unstable against nucleating virtual Alice
  string-loops on its surface, which will eventually destroy it
 even if we hold the string fixed. But the rate of this nucleation
 is strongly suppressed for even a modest hierarchy between the
 string-forming and $C$-breaking scales\ \lf\preswise\rr, so this is
  not a problem.}. \par
   As the temperature decreases, we assume that all three $a_\Is$
   parameters become small enough so that the $C$-symmetric vacuum
 $\vev{v^\Isp}=0$ becomes unstable. Let us begin by considering a single
Higgs triplet, say $I=1$. Since we would first like to understand
 the structure of the wall itself, without ambient matter, the
   $\psi$ fields will be set to zero at this stage. The relevant
 single-generation potential is then:
$$V_{wall}=V_1(\Phi)+\lm v^{\scriptscriptstyle T}\Phi v
+av^2+\rho(v^2)^2\eqno(2.9)$$
  In addition, the Higgs mechanism causes the two
 charged gauge bosons, $W_\pm$, to eat two of the components of
$\Phi$, and acquire a mass
$$M_{\scriptscriptstyle W}=3\sqrt{2}\al e\eqno(2.10)$$
 The VEV $\al$ defines the scale of the Alice {\it strings}.
 \par The $v=0$ vacuum is unstable, but $\vev{v_a}$ cannot be uniform
 throughout space, due to the Alice twist. Since $v_1$,$v_3$ are
  twisted
 by $-1$ around a closed curve linking the string, each of these two
 components must vanish on some surface having the string as boundary.
  As explained above, however,
  of the three isospin components only $v_3$ will develop
 a kink, i.e. a soliton configuration interpolating between two
 VEVs of opposite sign. The deviations of $v_a$ from their bulk
 VEVs will be limited to a wall, enclosing the $v_3=0$ surface and
  having a thickness of order
 $$ wall\; thickness\;=\; O(1/\sqrt{a})\,.\eqno(2.11)$$
 This is precisely the Alice wall (\fig{(1a) The Alice wall. (1b)
 An edge-on, cross-sectional view of the wall. The small circles
  represent sections of the Alice-string core.}).
  We let the string lie along a fixed circle in the $z=0$ plane; the
 $v_3=0$ surface is then taken to be the $z=0$ disk bounded by this
 circle.
  The $v$-field configuration of the wall
  reacts back, through the equations of motion, on the
   $\{\Phi,A_\mu^a\}$
 fields of its preexisting string boundary. Our approach shall be to
 choose the parameters so as to make this back-reaction small, and then
 to study the effect of the wall on a dilute ambient plasma,
   which in turn has negligible back-reaction on the Alice wall.\par
  We shall find it convenient to employ {\it two} distinct physical
 gauges in what follows; each of them has its advantages.
  One of them we call the `disk' gauge; it has its branch-cut inside the
 wall, at the surface where $\vev{v_3^{\Isp}}$ vanish\foot{As pointed
 out above, the inter-generation alignment potential $V_3$ makes it
 energetically favorable for the kinks of the three generations to
 be centered about the same surface}. For our static wall, this surface
 lies in the symmetry plane $z=0$. In this gauge,
 electric charge and baryon number are well defined outside
  the wall, and there is no difficulty in describing multi-wall
 configurations.
  The other physical gauge is the `transmission' gauge. Here
 the cut surface is chosen to extend from the string loop {\it outwards},
 to infinity, along the plane of the string. In this
  gauge charge and baryon number are not globally defined in the bulk,
  but it is easier to analyze baryon
 transmission since the cut lies away from the wall.
  The branch-cut surfaces for the two physical gauges are shown in
\fig{Edge-on view of the branch cut surfaces in two physical gauges:
(2a) disk gauge (2b) transmission gauge.}. \par
 In either of these physical gauges, the surviving components
 of the $\Phi$ field after the Higgs mechanism are:
 $$\Phi=
\left( \matrix{{\al+\varphi_1+\varphi_2} &\varphi_3 &0\cr
   \varphi_3 &{\al+\varphi_1-\varphi_2} &0 \cr
     0&0 &{-2\al-2\varphi_1}\cr} \right)  \eqno(2.12)$$
 Here $\varphi_1$ is neutral, whereas the two real fields $\varphi_2$,
 $\varphi_3$ together constitute a charged field. The masses of these
 $\Phi$ Higgs bosons are of order $\al$. Due to the assumed hierarchy,
 the wall thickness is much larger than the $1/\al$
 scale, and also larger than the related $W$-boson (inverse)
  mass scale: $$ a\ll\al^2,\quad a\ll e^2\al^2\eqno(2.13)$$
  We now require that the excitation of
 the $\Phi$-Higgs modes $\varphi_1$,$\varphi_2$,$\varphi_3$
 in the vicinity of the wall\footnote
 {\dag}{Actually, the neutral component
 $\varphi_1$ acquires a small VEV throughout space, but this
 serves only to modify $\al$ slightly.}
  be small relative to the bulk VEV,
$\al$. Upon consulting eqs.(2.9-13), we find this to hold for the
 following parameter range:
$$\lm(a/\rho)\ll\al^3\eqno(2.14)$$
Next, consider the back-reaction of the wall on the gauge fields;
 in the absence of the wall, $A_\mu^a$ vanish in a physical gauge.
 For a normal (non-superconducting) wall, it is easy to check that
 the gauge fields remain zero even near the wall. In the
  superconducting case, the gauge fields are excited in the wall,
   but again one can choose a regime where this back-reaction is
 small\ \lf\scond\rr. \par
 Having dealt with the issue of the wall's back-reaction on the
 Alice-string fields, it remains to study the wall configuration
 itself---that is, the $v$ kink. We find there are
 three pertinent Higgs-parameter ranges\ \lf\scond\rr: \par
$\bullet$ $a>2\lm\al$: the $v_a=0$ vacuum is stable, no wall forms.
 We call this regime I.\par
$\bullet$ Regime II: $-\zeta_0\lm\al<a<2\lm\al$. The
 $v=0$ vacuum is unstable, but a $v_3$ kink  with $v_1=v_2=0$ {\it is}
  stable. Here $\zeta_0$ is a pure number; we do not know it, but
  the relevant point is that $-2<\zeta_0\leq 11/2$,
   so regime II exists. The
  wall formed by this kink is normal (non-superconducting).\par
$\bullet$ Finally, regime III is $a<-\zeta_0\lm\al$. In it
 the above kink is unstable, and there exists another, stable kink
 with $v_2$ nonvanishing in the kink core and vanishing outside it.
 This kink is the superconducting wall referred to above.\par
   From here on we shall restrict attention to normal kinks.
 For our quartic potential, and in the {\it transmission} gauge,
  this kink is the well-known $tanh$ soliton centered at $z=0$
  \foot{This expression for the soliton requires corrections near
   the edge of the wall.}:
 $$v_1=v_2=0,\quad v_3=\sqrt{{{2\lm\al-a}\over{2\rho}}}\tanh(z\sqrt{
 2\lm\al-a})\eqno(2.15)$$
 Thus far, we have been discussing the structure of the Alice wall
 for a single Higgs triplet. When there are several
 generations, as in our model, we need to separately consider the
 three potentials $V_2^\Isp$. Assuming the temperature has fallen
 sufficiently so that all three $a_\Is$ have entered regime II, three
 kinks of type (2.15) will form, with different widths and heights;
 all are bounded by the same Alice string.
 As discussed above (2.7), the role of the inter-generation alignment
  potential $V_3$ is to break the degeneracy between the vacua
 with different relative signs of $\vev{v_3^\Isp}$, so that only two
 degenerate vacua are left. In these true vacua, all three VEVs
 have the same sign. Furthermore, since $C$ belongs to the original
 gauge group, these two vacua are gauge-equivalent and are thus
 physically identical. The spatial regions between kinks of different
 generations consist of false vacua, and the pressure differences between
 these regions and the true vacuum will force them to shrink. It is thus
 energetically favorable for the three kinks to coalesce. We therefore
 assume from here on that all three are centered at $z=0$
 \foot{Realistically their relative positions will fluctuate, at least
  for
 small $\eps$. However, they are more likely to be found near each other
 than far apart, so our assumption is reasonable even in a true
 finite-temperature environment.}
(\fig{Profiles of the three Higgs-triplet kinks.}).
 Finally, note that in addition to kinks bounded by the string, the
 $C$-breaking transition will also form kink bubbles with no boundaries.
 Since we are assuming a mass hierarchy, those
  bubbles which interpolate
 between the two equivalent vacua are metastable, while the other ones
 shrink and disappear due to pressure differences. The baryogenesis
 mechanism, to be demonstrated below, applies to the metastable bubbles
 as well as to Alice walls
 \footnote{\sharp}{As long as such a bubble does not begin
 to decay via virtual-string nucleation, however, a global definition of
  baryon number is not physically relevant. Once it decays
 it becomes an Alice wall, to which our analysis applies.}.
  According to refs.\wallnet ,
 however, bubbles tend to be shredded into bounded walls in an evolving
  string-wall network.
\chapter{The Spectrum and Discrete Symmetries}
 At temperatures below the $\vev{\Phi}$ scale, $\sod$ is
  broken down to $\ot$.
 The discrete symmetries of the vacuum are then $C$,$P$ and $T$.
The action of $C$ on the various fields is straightforward:
 for each gauge index $a\not=2$ appearing in a field,
  it gets multiplied
 by $(-1)$. Parity simply reverses the sign of spatial coordinates, and
 multiplies a field by $(-1)$ for every spatial index appearing in it.
 Time-reversal reverses the sign of time, changes c-numbers into their
 complex conjugates\footnote{\dag}{As mentioned above the baryogenesis
 mechanism explicated here is classical, although we choose
 to interpret it in particle language. At the classical level,
 then, one may work with the real components of all fields, and then
  no complex numbers are involved.}
  and, as for $C$ and $P$, multiplies
 the various fields by some signs. These signs are {\it almost}
 determined by the requirements that all couplings in the action
 are $C$,$P$ and $T$ invariant, and that $CT$ remains a symmetry when
 the Alice wall forms. The only freedom left in choosing these
 signs reflects the following two symmetries:
 \itemitem{(I)} $\psi\rightarrow-\psi$, other fields
  unchanged; this symmetry is just
    baryon-number conservation modulo 2, if perturbative
     baryon-violating processes are neglected.
\itemitem{(II)} A global $\uem$ rotation by $\pi$.\par
 Modulo (I) and (II), one finds that $T$ acts on the real fields
  precisely as $C$ does,
  except that $t\rightarrow-t$. In other words, $CT$ (which must be
 conserved by the $CPT$ theorem, since $P$ is) acts as follows in our
 model:
$$(\calc\calt)F(\vec{x},t)(\calc\calt)^{-1}=F(\vec{x},-t)
\eqno(3.1)$$
for any field $F$.
 When $v^\Isp$ acquire VEVs, $C$ and $T$ are broken
  as separate symmetries.
 Analogously to what happens for $CP$ violation in the Standard Model,
 the mass terms by themselves obey a modified $C$, but the interaction
 terms violate it, so in fact {\it no} good charge-conjugation symmetry
 can be found once the wall has formed. However, $CT$ remains a good
 symmetry, since the VEVs are left invariant under it (eq.(3.1)).
\par
Next, let us take stock of the spectrum after the formation of
 the Alice wall. The surviving components of the $\Phi$
 Higgs are $\varphi_1$,$\varphi_2$ and $\varphi_3$, which make up one real
 neutral field and one complex, charged field. Their
  masses are of order $\al$, if we make the quartic couplings in
 $V_1(\Phi)$ of order $1$.
  The charged gauge bosons $W_\pm$ have a mass given by (2.10),
 and the three neutral components of the $\psi^\Isp$ triplets also have
 masses $O(\al)$ (see (2.7-8)). We shall refer to these particles (with
 masses of order $\al$) as {\it superheavy}; they will decouple before
the wall-forming epoch, so we henceforth ignore them and the quantum
 processes they mediate.  \par
 The remaining particles are the charged $\psi$-particles and the
 (charged and neutral) $v$-particles. Since we
  wish to study the regime II
 (in the classification above (2.15)), we must choose
  $\lm_\Is$ small enough so that $$\lm_\Is=O(a_\Is/\al)\,.\eqno(3.2)$$
  We shall collectively denote the scales $a_\Is$ by $a$ where
 convenient. \par
  From (2.2),(3.2) we find that each generation of Higgs triplet
 has one real neutral field and one complex charged field
 \footnote{*}{We
 assume that the coupling $\eps$ is small, so the inter-generation
 mixing it engenders is negligible.} with masses
 of order $\sqrt{a}$:
$$(m_v^n)^\Isp=O(\sqrt{a}),\;(m_v^{ch})^\Isp=O(\sqrt{a})\eqno(3.3)$$
where the superscripts $n$,$ch$ refer to neutral and charged components,
 respectively. \par
Finally there are the three charged $\psi$ fields, which describe the
 three baryons and their antibaryons. Before $C$ breaking their masses
 were $m_\As$ (eqs.(2.7-8)). When $v^\Isp$ acquire VEVs,
 the Yukawa potential $V_5$ modifies the $\psi$ mass-squared
  matrix. The new
 mass term for the baryons is $z$-dependent; neglecting wall-edge
  effects, the baryonic action in the transmission gauge is
 $$S_{baryonic}=\int d^4x\bigl(\sum_\Is
 \prt_\mu\psi_-^\Isp\prt^\mu\psi_+^\Isp
-\sum_{\Is\Js}\psi_-^\Isp M^2_{\Is\Js}(z)\psi_+^\Jsp\bigr)\eqno(3.4)$$
where
$$\psi_\pm^\Isp\equiv{1\over{\sqrt{2}}}(\psi^\Isp_1\pm i\psi^\Isp_2)
\eqno(3.5)$$
are the (anti)baryon fields, and the
 position-dependent, hermitian mass-squared matrix for the $\psi_+
  ^\Isp$ fields is:
$$M^2_{\Is\Js}(z)=\tilde\nu_{\Is\Js}-ib\sum_\Ks\eps_{\Is\Js\Ks}
v_3^\Ksp(z)\eqno(3.6)$$
where $\tilde\nu_{\Is\Js}\equiv\nu_{\Is\Js}-\kp\al\delta_{\Is\Js}$ has
eigenvalues $m^2_\As$ (eqs.(2.4),(2.7)).
 The $v_3^\Isp$ kinks appearing in (3.6) are given by (2.15), with
 $(\lm,a,\rho)$ replaced by $(\lm_\Is,a_\Is,\rho_\Is)$.
  It is crucial for the baryogenesis
 effect that the widths of the three kinks all be distinct from
 one another, as will be seen in section 4. \par
  Let us denote
 $$r_\Is\equiv v_3^\Isp(+\infty)\eqno(3.7)$$
 These three numbers are positive
 \footnote{**}{Their relative signs are determined
 by the potential $V_3$, as discussed in section 2. Their absolute sign,
 however, can be changed by a global $C$ transformation, to which the
 physics is invariant.}, and determine the
  mass-squared matrix in the bulk.
 Switching to the disk gauge and choosing it so that $v_3^\Isp=r_\Is$
 away from the wall, we obtain from (3.6) the mass-squared matrix in the
 bulk:
$$ (M^2_{\Is\Js})_{\rm bulk}
=\tilde\nu_{\Is\Js}-ib\sum_\Ks\eps_{\Is\Js\Ks}
r_\Ks\quad (disk\; gauge)\eqno(3.8)$$
At this point, it is useful
  to change basis in generation space. We do this by
 acting with an $\sod$ matrix on the $I$ index, in such a way
that in the {\it new} basis (with indices denoted by
 the letters $A$,$B$, etc.)
$$(M^2_{\As\Bs})_{\rm bulk}=(\tilde\nu+\omega t_3)_{\As\Bs}\eqno(3.9)$$
where: $$\om=+b\sqrt{{\scriptstyle\sum}_\Is(r_\Is)^2}\,,\eqno(3.10)$$
 $$(t_\Cs)_{\As\Bs}=-i\eps_{\As\Bs\Cs}\eqno(3.11)$$
 are the generators of the generation-space rotation group,
and $\tilde\nu_{\As\Bs}$ is the matrix $\tilde\nu$ in the new basis.
 We now conveniently choose $\tilde\nu$ as follows:
$$\tilde\nu=m_h^2+(\om+m_l^2-m_h^2)(t_3)^2\eqno(3.12)$$
It follows from (3.9),(3.12) that the mass-squared spectrum of
baryons in the bulk is: $m_l^2$,$m_h^2$ and
$$m_\Hs^2\equiv m_l^2+2\om\,,\eqno(3.13)$$
and we choose $m_\Hs>m_h>m_l$.
Each baryon is degenerate with its antibaryon, as required by $CPT$
 invariance. The `light' ($l$) and `medium heavy' ($h$) baryons in
 the bulk are described (in disk gauge) by the $t_3=-1$ and $t_3=0$
 components of the $\psi_+$ field, respectively, whereas the `heavy'
 ($H$) baryon is described by the $t_3=+1$ component.
Thus, upon decomposing the baryon field in the $t_3$ basis,
$$\psi_\pm^\Asp={1\over{\sqrt{2}}}\sum_{\eta\in\{1,-1\}}
\psi_\pm^\etp\vector 1 {i\eta} 0 +
\psi_\pm^\zro\vector 0 0 1\;,\eqno(3.14)$$
we find the following disk-gauge field assignments in the bulk:
$$  baryons:\;(l,h,H)=(\psi_+^\mns,\psi_+^\zro,
 \psi_+^\pls)\eqno(3.15)$$
$$ antibaryons:\;
(\bar l,\bar h,\bar H)=(\psi_-^\pls,\psi_-^\zro,
 \psi_-^\mns)\eqno(3.16)$$ \par
 In what follows, we shall choose to decouple the $H$ baryon,
$$m_l<m_h\ll m_\Hs\eqno(3.17)$$
so that only the $l$ and $h$ baryons and their antibaryons
 will play a role in the plasma
 \footnote{\flat}{All three baryonic masses are
 still kept much lower than the superheavy scale.}.\par
So far we have discussed the mass-squared matrix in the bulk, and in
 disk gauge. Next, we return to (3.6) in the vicinity of the wall,
 and in the transmission gauge (in which it holds). The
  bulk assignments (3.15-16) now become:
$$\pm z>0\; baryons:\,(l,h,H)=(\psi_\pm^\mns,\psi_\pm^\zro,
 \psi_\pm^\pls)\eqno(3.18)$$
$$ \pm z>0\; antibaryons:\,(\bar l,\bar h,\bar H)=
(\psi_\mp^\pls,\psi_\mp^\zro,\psi_\mp^\mns)\eqno(3.19)$$ \par
It is important to note that although eqs.(3.18-19) hold in
 transmission gauge, our usage of the labels `baryons', `antibaryons',
  $l$,$\bar l$ etc. still
  refers to the {\it global} (disk gauge) definition.
 Thus an $l$ baryon impinging on the Alice wall from the $z<0$ side,
 for instance, may be transmitted through the wall to the $z>0$ side.
 In this case, the transmission-gauge description is that the $\psi_-$
 field is transmitted from a $t_3=-1$ to a $t_3=+1$ component,
 and the outgoing particle is called $\bar l$. In the disk gauge,
 however, we would say that the $\psi_+^\mns$ field ($l$) acquires
 upon transmission a $\psi_-^\pls$ component ($\bar l$). \par
 The mass-squared matrix throughout space, including the wall core, is
 $$M^2(z)=\tilde\nu
 +\sum_\As\omega_\As(z)t_\As\quad\,(transmission\;gauge)
 \eqno(3.20)$$
 where $\omega_\As(z)$ are related to $v_3^\Isp$ by the $SO(3)$ matrix
which relates the two generation-space bases. The functions $\om_1$,
 $\om_2$ are thus
only defined modulo an arbitrary $SO(2)$ rotation, but the physics we
shall be interested in is unaffected by this ambiguity.
 The functions $\om_\As(z)$ satisfy (see (3.9))
$$\omega_3(\pm\infty)=\pm\omega\; ,\eqno(3.21a)$$
$$\omega_\As(\pm\infty)=0\quad{\rm for}\; A=1,2\; .\eqno(3.21b)$$
All the information on transmission and reflection of baryons and
 antibaryons at the wall is encoded in the following Klein-Gordon
 equation ((3.4),(3.20)):
 $$\{\prt^2+\tilde\nu+\sum_\As\om_\As(z)t_\As\}\psi_+=0\eqno(3.22)$$
 with $\tilde\nu$ given by (3.12).
\chapter{Baryogenesis via Asymmetric Transmission}
Let us consider a temperature low enough so that the Alice wall has
 formed. The wall is embedded in a plasma of baryons and antibaryons
(and other particles) having an initially-zero net baryon number,
and we are interested in baryon-number violating processes capable of
giving rise to baryogenesis.\par In the {\it bulk} of the plasma
one has the usual kinds of baryon-violating processes, familiar from
realistic GUTs. There are perturbative quantum processes, all of them
suppressed by inverse powers of the superheavy scale; A few such
processes are depicted in \fig{Some perturbative
  baryon-violating processes. Diagrams (4a-c) are
   suppressed by superheavy
 propagators. The plasma contains a negligible $W_\pm$ population, so
 (4d) can be ignored as well.} (In general one also expects
 sphaleron-induced baryogenesis, but that requires chiral fermions
 and hence cannot arise in our model).
 The wall-catalyzed baryogenesis mechanism, however, is independent
 of the superheavy scale\foot{Except in that the Alice string itself was
 formed at that scale.}. Furthermore, we shall see that
 it can be catalyzed even by a static wall, if the wall disappears at
 finite time. We now investigate this mechanism in detail, for a single
 static wall.\par In order to facilitate a clear
presentation, the remainder of this section is divided into subsections
 mirroring its main conceptual components, which are:
 \itemitem{1)} Choice of a convenient {\it parameter regime}.
 \itemitem{2)} The {\it wall-catalyzed baryonic $S$-matrix}; $CPT$
 constraints.
 \itemitem{3)} {\it Averaged rate equations}: approximate, linear
 Boltzmann equations for the plasma populations of the particle species
 $\{l,h,\bar l,\bar h\}$ in the presence of the wall, averaged over
 particle positions and momenta.
 \itemitem{4)} {\it Calculation of the microscopic baryon-antibaryon
 asymmetries} (reflection and transmission), in a Born approximation.
 \itemitem{5)} {\it Demonstration of finite-time
  baryogenesis} $B(t)$ for
a plasma which, at $t=0$, is out of thermal equilibrium and has $B(0)=0$.
 \itemitem{6)} {\it Different kinds of Alice walls}.
\smallskip {\bf 4.1 Parameter Regime.--}We choose the range of model
 parameters and plasma temperature so as to simplify the analysis and
 suppress uninteresting effects. Let $m_u$ be the mass of the (charged
  and neutral)
  $u$-bosons, which are implicitly present in our model (see discussion
 above (2.1)). Let $T$ be the temperature, and define $\eta_\Is\equiv
2\lm_\Is\al-a_\Is>0$. The regime we work in is defined in the appendix.
It implies the following:
\itemitem{(I)} The wall is normal (not superconducting).
\itemitem{(II)} The plasma is populated mainly by nonbaryonic
$u$-particles, and its baryonic
 component is dilute and non-relativistic.
The back-reaction of the plasma upon the wall is negligible.
\itemitem{(III)} The populations of $H$,$\bar H$ baryons in the plasma,
as well as those of the $v$-particles and the superheavies,
 can be approximated as zero.
\itemitem{(IV)} A baryon ($l$ or $h$) can only disappear by annihilating
 an antibaryon. Such annihilations occur at negligible rates.
In general, since $u$-particles are far more abundant than baryons
we may ignore reactions involving more than one baryon (or antibaryon)
in the initial state. Collisions (predominantly electromagnetic)
 between baryonic particles and
 $u$-particles serve to equilibrate the momenta of the former, but the
probability of such a collision changing the {\it species} of a baryonic
particle (from $l$ to $h$, etc.) is suppressed\foot{Because gauge
 interactions are diagonal in generation space.}.
\itemitem{(V)} The $\om_1$,$\om_2$ terms in (3.22) are small enough
 to be treated in a Born approximation (but not $\om_3$); this will
 prove useful in subsection 4.4.
\smallskip {\bf 4.2 Wall Catalyzed $S$-Matrix.--}
The plasma particles exert a negligible back-reaction on the Alice wall;
hence they can exchange momentum, but not energy, with it. The
 creation of baryon-antibaryon
pairs at the wall is thus kinematically suppressed
 \foot{We shall ignore electromagnetic bremstrahlung at the wall,
 since it does not change the (anti)baryon species (see (IV) above).}.
  Consulting point (IV), we see that
the only species-changing
microscopic baryonic processes occurring in
the plasma are the reflection or transmission of an $l$,$\bar l$,$h$ or
$\bar h$ particle at the Alice wall. Therefore the baryonic rate
 (or Boltzmann) equations, to be considered in subsection 4.3, are
linear in the concentrations of these four particle species and
  determined by the reflection and transmission rates. Since
the normal Alice wall considered here has no charged zero modes,
impinging particles cannot exchange localized charge with it. The
{\it wall-catalyzed $S$-matrix} thus has nonzero entries only for the
 following processes:
\itemitem{$\bullet$} $\Delta B=0$ reflection:
 $S_{ij}$, with $i$ the
initial-particle label, $j$ the final; $i,
j\in\{l,\bar l,h,
\bar h\}$; and $(i,j)$ both baryons or both antibaryons.
\itemitem{$\bullet$} $\Delta B=\pm 2$ transmission:
 $S_{ij}$, with
$i\in\{l,h\}$; $j\in\{\bar l,\bar h\}$ or vice versa.
For transmission processes, the wall picks up an electric Cheshire charge
$Q=-\Delta B$. Some reflection and transmission processes are depicted in
 \fig{Some reflection and transmission processes in the presence
 of the Alice wall.}.
 \par The $S$-matrix depends on the incoming momentum, but the
{\it outgoing} ($j$-particle) momentum is uniquely determined by
 conservation of energy
  and $\vec{k_\prl}$ (momentum parallel to wall):
$$S_{ij}=S_{ij}(\vec{k}_i)\,.\eqno(4.2.1)$$
As before we are ignoring wall-edge effects, which would slightly
 violate $\vec{k}_\prl$ conservation. For particles that miss the wall
 altogether $S_{ij}$ is, of course, the unit matrix
  $\delta_{ij}$. For kinematical reasons,
 $$S_{ij}(\vec{k})
 =0\quad\;{\rm if}\;\;\vec{k}^2<m_j^2-m_i^2\,. \eqno(4.2.2)$$
Since $\vec{k}_i$ determines $\vec{k}_j$, unitarity
 simply means that the $2\times 2$ complex matrix
  $S_{ij}(\vec{k})$ is unitary. The wall
is physically symmetric under $z\rightarrow -z$, as is manifest in
 disk gauge; thus $S_{ij}(\vec{k})$ does not depend
  on the sign of
$k_z$---the particle $i$ may be incident on either side of the wall.
 The $CPT$ symmetry implies that
$$S_{ij}(\vec{k}_i)=S_{\bar\jmath\bar\imath}(\vec{k}_j)
\eqno(4.2.3a)$$
when $$E_i(\vec{k}_i)=E_j(\vec{k}_j),
\quad\vec{k}_\prl^{(i)}=
\vec{k}_\prl^{(j)}\,.\eqno(4.2.3b)$$
Here $E_i(\vec{k})\equiv\sqrt{m_i^2+\vec{k}^2}$.\par
This constraint relates, for instance, $S_{ll}(\vec{k})$ to $S_
{\bar l\bar l}(\vec{k})$. Thus, had we chosen a parameter range such
 that only
the lightest baryon is present in the plasma, no baryogenesis could
 occur\foot{We thank J. Preskill for a discussion on this point.}.
However, since we have $h$-particles as well, the $CPT$ symmetry cannot
 rule out asymmetries enabling the production of a net baryon number.
\smallskip {\bf 4.3 Averaged Rate Equations}.-- In order to
simplify the four-species Boltzmann equations into a set of four
first-order, linear, ordinary differential equations, we assume that the
initial plasma has Boltzmann distributions in all particle energies---
 but only within each species. In other words, denoting by
  $N_i(\vec{k},
t)d^3k$ the momentum-space distribution of particle momenta at time $t$,
we assume
$$N_i(\vec{k},0)=n_i(0)\bolt/\int d^3k\bolt\eqno(4.3.1)$$
where $n_i(t)$ is the total number of $i$-particles. The exact
rate equations do {\it not} preserve this Boltzmann momentum distribution
 within a given species, unless $n_i$ are also in the correct
  Boltzmann
ratios at $t=0$. However, the baryonic particles in the plasma are
 constantly undergoing electromagnetic collisions
  with $u_\pm$ particles, and this
tends to restore their Boltzmann $\vec{k}$ distribution {\it without}
affecting the ratios $n_i/n_j$ (point (IV)).
 Since the mean free path between such
collisions is much shorter than a baryon's mean path between consecutive
  encounters with the wall, we will assume that the momentum distribution
 within a species stays frozen at its initial, Boltzmann shape
 \foot{The baryonic plasma
 is dilute, so the Bose-Einstein and Boltzmann distributions agree.}
at all times $t>0$. However, the total population of the
 $i$-th species,
 $n_i(t)$, does evolve with time.
The non-equilibrium nature of the initial plasma, necessary for
 baryogenesis, will then manifest itself only through non-Boltzmann
initial
 population ratios, $n_h(0)/n_l(0)$ and $n_{\bar h}(0)/n_{\bar l}(0)$
\foot{These ratios could have realistically deviated from equilibrium
  as $T$ first decreased through $m_h$ and then
through $m_l$.}. Since the wall
interconverts light and medium-heavy particles, the rate equations will
 tend to restore these ratios to their Boltzmann values.\par
A note on spatial dependence: the generated baryon number will actually
diffuse outwards from the Alice wall; we assume
 a very large diffusion coefficient, so the species populations
  $n_i(t)$
can be taken to be spatially homogeneous. We have not attempted to
translate this simplifying
 assumption into an additional quantitative condition on
 our model parameters. In our thought experiment,
 we also stipulate that the wall is formed (or externally introduced)
at $t=0$ into a plasma box of finite volume $V$. This is reasonable:
in a cosmological setting, a gas of Alice walls will form, so $V$ may
 be thought of as the average volume per wall.\par Given a plasma
 with such properties, then, we
  now derive the averaged (over space and momenta) rate
 equation satisfied by $n_i(t)$. Let $A$ denote the area of
  the wall,
which is assumed much larger than all baryonic Compton wavelengths;
  we are then justified in neglecting wall-edge effects. We define
 $P_{ij}$ to be the probability per unit time
  that a given particle of species $i$, anywhere in the box,
 is converted by the wall into a $j$-particle. We have:
$$f_i P_{ij}={A\over V}\int {{\vert k_z\vert}
\over{E_i(\vec{k})}}
d^3k\bolt\vert S_{ij}(\vec{k})\vert^2\,,\quad j\not=
i\eqno(4.3.2)$$
where:$$f_i=f_{\bar\imath}\equiv\int d^3k\bolt\,.\eqno(4.3.3)$$
In (4.3.2), $\vert k_z\vert/E_i$ is the normal velocity of the incoming
 baryon or antibaryon relative to the wall, and no summation over $i$
 is implied. By $S$-matrix unitarity,
  $$f_i\sum_{j\not=i}P_{ij}={A\over V}\int{{
  \vert k_z\vert}\over{E_i(\vec{k})}}d^3k\bolt(1-\vert
S_{ii}(\vec{k})\vert^2)\,.\eqno(4.3.4)$$
The $CPT$ invariance yields the relation $$f_i P_{ij}=
f_{\bar\jmath}P_{\bar\jmath\bar\imath}\,.\eqno(4.3.5)$$ where we have
 used the fact that, for a transition $i\rightarrow j$,
$$\vert k_z^{(i)}\vert d^3k_i=\vert k_z^
{(j)}\vert d^3k_j\eqno(4.3.6)$$
This follows from the conservation laws (4.2.3b) and is simply a
 manifestation of the conservation of phase-space volume in
 hamiltonian mechanics (Liouville's theorem).\par
 The averaged rate equations now assume the form, familiar from
 statistical mechanics:
$${{dn_j(t)}\over{dt}}=\sum_{i\not=j}
 n_i P_{ij}-n_j\sum_{i\not=j}P_{j
 i}\quad\quad i,j\in\{l,h,\bar l,\bar h\}\,.\eqno(4.3.7)$$
\smallskip {\bf 4.4 Calculation of Microscopic Asymmetries}.--
As noted at the end of section 3, the wall-catalyzed
 $S_{ij}$ is encoded in the Klein-Gordon equation
(3.22). We can work with this single equation, and
 yet describe (in transmission gauge) both baryons and antibaryons. This
 can be seen from the assignments (3.18-19), which tell us that
 eq.(3.22) describes baryons for $z>0$ and antibaryons for $z<0$.\par
 We shall assume the $\psi_+$ wave is monochromatic, so the
 methods of time-independent scattering theory can be used.\par
 The idea of the Born approximation we shall employ, is to perturb
 $\psi_+$ (to first order in $\om_1$,$\om_2$) around the solution
 of (3.22) with only the $A=3$ term. The parameter regime chosen in
 4.1 ensures that this Born approximation is valid, and that the limit
$m_\Hs\rightarrow\infty$ may be taken with the other two baryon
masses remaining finite.
The truncated space of wave functions (`Hilbert space' in a
first-quantized picture) consists, in this limit, of functions with
only the two components $t_3=-{\rm sgn}z,0$ at position $(x,y,z)$.
 In addition, the $t_3\not=0$ components vanish at $z=0$, whereas the
 $t_3=0$ component and its first derivative are continuous there.\par
The parameter range further ensures that the width of the $I=3$
kink is very small compared with those of $I=1,2$,
 and these latter widths
 are in turn much smaller that the de Broglie and Compton wavelengths
 of the baryons. The unperturbed $\psi_+$ wave for an incoming $l$
 is thus the wave function of a plane wave, impinging on a `brick wall'
 (at $z=0$) from the $z>0$ side and totally reflecting from it.
  It is therefore the sum of the incident and
reflected waves. The unperturbed $\psi_+$ wave for an incoming $\bar l$,
 on the other hand, is the sum of a plane wave incident from
 the $z<0$ side, and a totally-reflected wave.
  For an incoming $h$ or $\bar h$, the potential is flat and the
  zeroth-order $\psi_+$ is just the incoming plane wave, from the
 appropriate direction.\par
The perturbing potentials $\om_1(z)$,$\om_2(z)$ are finite in the
$m_\Hs\rightarrow\infty$ limit, and are given by eq.(A.10) of the
 appendix. They are also depicted in \fig{The potentials of the
 Klein-Gordon eq.(3.22) in the $m_\Hs\rightarrow\infty$ limit.
 In this limit, the amplitude of the
 $I=3$ kink of figure 3 becomes infinitely large, and this kink
  becomes infinitely narrow. The two solid curves
are profiles of $\om_1(z)$ and $\om_2(z)$, while the dashed horizontal
lines are the two surviving mass-squared levels.}. \par
Using the zeroth-order waves described above, it is straightforward
 to compute the various entries in the wall-catalyzed $S$-matrix.
 For instance, the element $S_{lh}$ in the Born approximation reads
 up to a phase (where kinematically allowed by (4.2.2))\foot{For
  typical (thermal) momenta, $\vert S_{lh}\vert^2\ll 1$ thanks to
 (A.7). For the kinematical endpoint $k^\pr_z\rightarrow 0$, the
 expression (4.4.2) is singular. This is an artifact of the Born
 approximation, and has a negligible effect on the momentum-averaged
 asymmetry ((4.4.4a) below).}
$$S_{lh}(\vec{k})=S_{\bar h\bar l}(\vec{k}^\pr)\approx{1\over{\sqrt{
\vert k_z\vert\vert k^\pr_z\vert}}}\int_0^\infty dz e^{-i\vert k_z^\pr
\vert z}\sin{(\vert k_z\vert z)}\langle 0\vert\sum_{\As=1}^2\om_\As
(z)t_\As\vert -1\rangle\,,\eqno(4.4.2)$$
where $\vec{k}=\vec{k}_l$, $\vec{k}^\pr=\vec{k}_h$ and $\vert
\zeta\rangle$
 is a generation-space state having $t_3\vert\zeta\rangle=\zeta\vert
\zeta\rangle$. As
will be seen in 4.5, all the relevant baryon-antibaryon transmission
asymmetries are expressible in terms of a single {\it reflection}
asymmetry, by use of the $CPT$ constraints and unitarity. This
asymmetry is
$$\cala_{hl}\equiv P_{\bar h\bar l}-P_{hl}\,,\eqno(4.4.3)$$
And in the Born approximation it is:
$$\cala_{hl}\approx -2{A\over V}\vev{{1\over{\vert k_z^\pr
\vert E_h(k)}}{\rm Im}\{D_1(k)D^*_2(k)\}}\,,\eqno(4.4.4a)$$
where now $\vec{k}$ is the incoming ($h$ or $\bar h$)
 momentum, $\vec{k}^\pr$
the outgoing ($l$ or $\bar l$) momentum, the average is performed over
 incoming momenta with Boltzmann measure, and the complex amplitudes
 $D_\As$ are defined as follows:
$$D_\As(k)\equiv\int_0^\infty dz e^{-i\vert k_z\vert z}\sin
(\vert k^\pr_z\vert z)\om_\As(z)\eqno(4.4.4b)$$
with $\om_\As$ given by eq.(A.10).
 This asymmetry is, in general, nonzero. One easily checks
 that the minimal numbers of baryonic and Higgs generations for which
 $\cala_{hl}$ does not vanish identically, are three and three;
  this statement holds to all orders in perturbation theory.\par
 In subsection 4.5 we will show that the plasma, which starts at $B(0)=0$
and out of equilibrium, develops a finite $B(t)$ thanks to the asymmetry
 $\cala_{hl}$. Note that this baryogenesis results even though we
 chose to decouple one of the three baryon mass-eigenstates
  (by making its mass $m_\Hs$ high). This does not contradict our above
statement, which implied that two baryonic generations cannot yield
baryogenesis. This is because, of the surviving baryons and antibaryons,
$h$ and $\bar h$ are in the same gauge multiplet, {\it but
 $l$ and $\bar l$ are not}--- they are not related
 by $C$ (see (3.15-16)). The dynamics of the Alice wall still involves
all three baryon generations, even when $H$ is decoupled.
\smallskip {\bf 4.5 Finite-Time Baryogenesis}.-- From unitarity and
$CPT$-invariance (eqs.(4.3.3-5)) it can be shown that the rate equation,
(4.3.7), admits a unique equilibrium state:
$$(n_i)_{\rm equil}\propto f_i\,.\eqno(4.5.1)$$
Now in general, the net plasma baryon number is
$$B(t)=n_l+n_h-n_{\bar l}-n_{\bar h}\eqno(4.5.2)$$
so from (4.5.1), $B_{\rm equil}=0$; by assumption, the initial baryon
number is also zero, $B(0)=0$. However, at finite $t$, $B(t)\not=0$,
provided the initial ratio $n_h/n_l$ differs from the equilibrium
ratio $f_h/f_l$. To see this, let us compute $\dot B(0)$. From (4.3.2-7),
$$\dot B(0)=-2\cala_{hl}(n_h-f_hn_l/f_l)\,,\eqno(4.5.3)$$
where $\cala_{hl}$ is the asymmetry computed in subsection 4.4.
If the wall remains in place, then $B(t)$ must eventually approach zero
again. If the wall disappears at some finite time, however, then a
baryon asymmetry will be left behind. \par
In general, (4.3.7) is solved by rescaling $n_i=\sqrt{f_i}\tilde n_i$
 and expanding the initial distribution in
the eigenvectors of $\tilde P_{ij}-\delta_{ij}q_i$, with
 $\tilde P_{ij}=\sqrt{f_j}P_{ji}/\sqrt{f_i}$, $q_i\equiv
\sum_{k}P_{ik}$ and $P_{ii}\equiv 0$. Since
 $\tilde P$ is real and asymmetric\foot{Its asymmetry is solely due
 to reflection and transmission asymmetries, such as $\cala_{hl}$.},
 the corresponding eigenvalues are pairs of complex-conjugate complex
numbers. Thus $n_i(t)$, and therefore $B(t)$ as well,
 are linear combinations of exponentially
damped sine functions with various periods, phases and decay constants.
 This asymptotic behavior of $B(t)$ (for a wall held in place
indefinitely) is more general than the particular model considered here.
 To see a particularly elegant
 example of this behavior, let us concoct a simple-minded $P$ matrix
which realizes the necessary baryon-antibaryon asymmetry in a `maximal'
way. The $P$ matrix we choose is deterministic---the label
 of the incoming
particle determines that of the outgoing one, and thus whether
 the incoming
particle is reflected or transmitted. The rules we choose are
 encapsulated in \fig{A set of
 deterministic reflection/transmission rules with maximal transmission
 asymmetry.}: an $l$ is always reflected (into an $h$); an $\bar h$
 reflects into an $\bar l$; an
 $\bar l$ is always transmitted into an $l$,
 whereas $h$ is always transmitted to become an $\bar h$.
 It is readily seen that the asymmetry is then
 nonzero. For simplicity we set $m_h=m_l$, so that all nonzero
 entries in $P_{ij}$ are equal to $\tau^{-1}\equiv{A\over V}
 \vev{\vert v_z\vert}$,
 with $\vev{\vert v_z\vert}$ the average $z$-velocity of any incident
particle. It is then straightforward to show that, as long as the wall
 is present, $$B(t)=-2(n_h(0)-n_l(0))e^{-t/\tau}\sin{(t/\tau)}\,.\eqno
 (4.5.4)$$
\smallskip {\bf 4.6 Different Kinds of Walls.--}
As discussed in section 2, the inter-generation alignment term $V_3$
correlates the relative signs of the three bulk VEVs $\vev{v_3^\Isp}$,
and also causes the three types of kinks to be preferentially centered
at the same
surface. This is important for baryogenesis: if the alignment coupling
 $\eps$ is set to zero there are degenerate, nonequivalent vacua with,
say, the same VEV sign for $I=1,3$, but opposite signs of
 $\vev{v_3^{(2)}}$. Therefore, by eqs.(3.7),(4.4.4) and (A.10), Alice
  walls embedded in these two vacua will yield different
 asymmetries $\cala_{hl}$ --- in fact, in the decoupled-$H$ limit the
  asymmetries are equal and opposite.
 When $\eps\not=0$, however, the bubbles enclosing one
 type of vacuum tend to collapse due to the pressure difference between
 the vacua, and so only one $\cala_{hl}$ value is relevant.\par
 For topological and dynamical reasons,
 there is only one kind of Alice wall
 (bounded by one type of Alice string) in the true vacuum, since
 an Alice wall must be a solution of the equations of motion. However,
  suppose our plasma
 is inhabited by capable engineers who are able to change the
  {\it shapes} of
  the $v_3^\Isp$ kinks, without changing their VEVs
in the bulk of the plasma. Assume further that they are able to
 hold these altered kinks static by some means. Such an
 {\it artificial wall} is still bounded by an Alice string.
 If an artificial wall is adjusted so that its $\om_1$,$\om_2$ functions
 are interchanged, we see from (4.4.4),(4.5.3) that the baryogenesis
 it causes will be precisely opposite to that engendered by
 a genuine (`natural') Alice wall of the same area. One can easily
show that this conclusion holds for the exact Klein-Gordon equation,
 and is thus not limited to the Born approximation.
 Fortunately, however, even the full transport equations are unlikely
 to give rise to engineers --- so our conclusions are safe.\par
Although there is only one kind of Alice wall (in the true vacuum),
 two otherwise identical Alice walls can
 still be different---they may carry different amounts of electromagnetic
 charges. More specifically, there are two such charges\ \lf\preskill
\rr : Cheshire electric charge and Cheshire magnetic charge.
The electric Cheshire charge (caused e.g. by transmission of baryons)
 decays rapidly, mostly by attracting oppositely charged $u$-particles
  from the plasma, which are almost always transmitted
 and thereby cancel the Cheshire charge. We
 have ignored magnetic charges in this paper;
 in ref.\dtmon\   a scenario
 was presented in which they disappear. But even if they are present,
 their effect on the wall $S$-matrix is not expected to
cancel the baryogenesis found here---especially since in a gas of Alice
walls, the sign of a given wall's magnetic charge is random.\par
\chapter{Summary and Conclusions}
In a Grand Unified Theory in which the charge conjugation operator is
 contained in the original gauge group, Alice strings form when the
 symmetry is broken to a smaller group having $C$ as a discrete symmetry.
 Magnetic monopoles usually form as well, but (as shown in ref.\dtmon)
  these can be eliminated by a Langacker-Pi mechanism, at
 least in a toy model. Thus magnetic monopole bounds need not rule out
 cosmological, post-inflationary Alice strings.\par In this paper we
 have studied, also in a toy model, a novel baryogenesis mechanism
 involving Alice walls. The original gauge group is $\sod$; baryons
 are scalar and belong to $\sod$ triplets. Baryon number $B$ is violated
 by perturbative processes, which become suppressed after $\sod$ is
 spontaneously broken to  $\ot$. However, this same transition forms
 Alice strings, in the presence of which $B$ is globally ill-defined.
 At a lower temperature, $C$ too is spontaneously broken, through VEVs of
 non-baryonic $\sod$ Higgs triplets. A closed Alice string loop now
 becomes the boundary of a domain wall---an  `Alice wall', where the
 Higgs triplets develop kinks. In a cosmological network of
 string-bounded walls, the walls eventually shrink and decay, and never
 dominate the energy density of the Universe; $C$ and $CP$ remain broken.
While the walls exist, the symmetries $C$, $CP$ and $B$ are all violated,
the latter mainly by the transmission of baryons/antibaryons
  through the wall. This enables baryogenesis, which indeed occurs
(even for static walls)
  when there are three baryonic `generations' and three Higgs triplets,
 and when the initial plasma (with $B=0$) is out of thermal equilibrium.
The baryogenesis occurs due to different transmission rates of baryons
 and antibaryons through the wall. Such a transmission changes a baryon
 into an antibaryon and vice versa, in the `disk' gauge where baryon
 number is well-defined in the region excluding the disk.
The Alice walls must be removed at finite time, or else the net baryon
number equilibrates back to zero.\par This baryogenesis mechanism is
 classical in nature. As baryon number increases the Alice wall
  tends to accumulate electric charge; in the parameter
 regime that we have focused on in this paper, this charge is purely
 of the `Cheshire' (unlocalized) variety, and decays rapidly,
 predominantly by attracting oppositely-charged particles from the
 ambient plasma and flipping their charges.
  These charge-decay processes have a negligible effect on the produced
  baryon number for the parameters we choose. They
 merely serve to electrically screen the net baryon excess by
 a corresponding excess of oppositely charged,
  non-baryonic particles.\par
The main conceptual difference between our mechanism and electroweak
 scale, sphaleron-induced baryogenesis\ \lf\dine\rr\ is as follows.
 Electroweak baryogenesis assumes a preexisting $CP$-violation, uses
 sphalerons to supply $B$-violation, and requires moving bubble-walls.
  The Alice-catalyzed
 mechanism, on the other hand, relies on preexisting deviations from
 equilibrium, and uses unstable domain walls to supply the other Sakharov
 conditions\foot{The phase transition that gives rise to Alice walls
 is also likely to increase the deviations from equilibrium, thus
  enhancing baryogenesis.}.\par
 Some realistic GUT schemes can form Alice strings and walls, and
these might catalyze cosmologically significant baryogenesis of the
 type found in the toy model.
 This possibility, for the unification group $\soten$, is currently
  under investigation in a scenario in which $C$,$P$ and $CP$
 are spontaneously broken at the left-right symmetry breaking
  scale. Both in the toy model and $\soten$ GUTs, topologically stable
  bubbles may form, in addition to Alice walls. In the toy
  model, such bubbles shrink away due to pressure differences and do not
 affect the wall-catalyzed baryogenesis. We do not yet know how
 non-Alice walls affect similar baryogenesis mechanisms for $\soten$,
 but the different nature of $C$ and $CP$ violations is likely to
 play an important role in that (or any other realistic) model. Finally,
 it remains to be seen to what extent the baryon number produced at the
 left-right symmetry breaking epoch can survive the above-mentioned
 electroweak-sphaleron processes.
\appendix
The constraints we choose to impose on the temperature and the
parameters of the toy model, are as follows (apart from the condition
that regime II of section 2 holds):
$$\lm_\Is=O(a/\al)\eqno(A.1)$$
$$\rho_\Is\gg a^2/\al^4\quad(I=1,2,3)\eqno(A.2)$$
$$\rho_3=O({{ab^2}\over{m_\Hs^4}}),\;\rho_\Is=O(1)\quad(I=1,2)
\eqno(A.3)$$
$$m_u\ll T\ll m_l<m_h\,,\eqno(A.4)$$
$$m_h\ll m_v^{(n,ch)}\ll m_\Hs\ll M_W\eqno(A.5)$$
$$(m_\Hs/\al)^4\ll b^2/a\eqno(A.6)$$
$$b^2\ll m_lT\eqno(A.7)$$
$$\sqrt{\eta_\Is}\ll\sqrt{a}/\ln(m_\Hs^4/(b^2\eta_\Is))
\quad(I=1,2)\eqno(A.8)$$
$$\eta_3=O(a)\eqno(A.9)$$
Condition (A.1) is eq.(3.2), and follows from the requirement that
the three kinks are normal (regime II of section 2). Condition (A.2)
follows from (2.14) (small back-reaction of the wall on the string)
together with (A.1). The choices (A.3-9) are consistent with
(2.15),(3.10),(3.13). Eq.(A.7) ensures the validity of the Born
 approximation in section 4.4, and (A.6) follows from (A.2-3).
\par
For the purposes of the wall $S$-matrix calculations of 4.4, the
 large-$m_\Hs$ limit results in the following approximation for the
 perturbing potentials in (3.22):
$$\om_\As(z)\approx br_\As(\tanh(\sqrt{\eta_\As}z)-{\rm sgn} z
)\,,\quad\,A=1,2\,.\eqno(A.10)$$
\ACK{We are indebted to
L. Dixon, P. Huet, H.P. Nilles, M. Peskin, J. Preskill and
 L. Susskind for useful discussions.}
%\endpage
\refout
\figout
\bye